\documentclass[article,twocolumn,amssymb,aps,prd,groupedaddress,superscriptaddress,nofootinbib]{revtex4-2}
\usepackage{amsmath, amssymb}
\usepackage{hyperref}
\usepackage{graphicx}
\usepackage{bm}
\usepackage{mathrsfs}
\usepackage{enumerate}
\usepackage{tensor}
\usepackage{color}
\usepackage{soul}

\begin{document}

\title{A Note on the Stability of the Dark Energy Model from Time Crystals}

\author{Laura Mersini-Houghton}
\email{mersini@physics.unc.edu}
\affiliation{Department of Physics and Astronomy, UNC-Chapel Hill, NC, USA 27599}

\date{\today}

\begin{abstract}
In this note, we investigate the stability of the dark energy model from time crystals proposed in ~\cite{laura1}. We emphasize two ingredients, the coupling of the scalar field to gravity, and the fact that these times crystals are on an expanding FRW background, which play a crucial role in the field's dynamics. The Hubble parameter, which contributes a drag term to the equations of motion, grows with time until the scale factor diverges. When taken into account, these factors also alleviate the stability concerns of ~\cite{cline}.  
\end{abstract}

\maketitle
\flushbottom



Time crystals spontaneously break time translation symmetry. 
They have been the subject of intense interest, since their discovery in ~\cite{Shapere2012,Wilczek2012}. Various mechanisms that produce time crystals have been proposed, ranging from non canonical scalar field models, to periodically driven Floquet quantum systems, disorder driven Anderson type localized waves, or two level systems, ~\cite{timecrystalreview}. An attractive feature they share is their stability against perturbations. 

In \cite{laura1} the author observed that the original proposed model of time crystals has an equation of state reminiscent of phantom dark energy due to its non canonical nature. However, in contrast to k-essence models where the phantom field can explore large regions of space, in the time crystals model the field is confined to a small region of field space within the boundaries of the crystal.

Non-canonical scalar field models have been extensively studied in cosmology due to their rich dynamics and potential to explain dark energy and inflationary phenomena~\cite{ArmendarizPicon2000,Chiba2000,ArmendarizPicon2001, Melchiorri_2003}. A common concern has been the possibility of ghost or gradient instabilities (see ~\cite{nima} and reference therein) in those models. As is well known, the coupling of the scalar fields to gravity in the expanding background of a Friedmann-Roberston-Walker (FRW) universe changes the stability requirements and often drives the field to an attractor point, ~\cite{nima, ArmendarizPicon2001, Chiba2000}. Here we explore the stability of the model introduced in ~\cite{laura1} in the background of an FRW universe.

The model presented in ~\cite{laura1} is described by the Lagrangian density
\begin{equation}
\mathcal{L} = g(X) - V(\phi), \label{eq:lagrangian_general}
\end{equation}

where \( \phi(t,x) = \phi_{0}(t) + \delta\phi(t,x) \) is a scalar field with a homogeneous background field \(\phi_0\), (as is typical of inflationary fields),  and perturbations around the background that depend on space and time. Here, \( X = \frac{1}{2} \nabla^{\mu}\phi \nabla_{\mu}\phi \) which for the homogeneous field simply becomes \( X = \frac{1}{2} \dot{\phi_0}^{2} > 0\) where the dot denotes a time derivative. \(V(\phi)\) is a positive semi-definite function of \( \phi \); \( g(X) \) is a non trivial function of \( X \) which for this model takes the form
\begin{equation}
g(X) = - \kappa X + \lambda X^2
\label{eq:g}
\end{equation}
where \(\kappa, \lambda\) are positive constants.

From now on, we will drop the index \(_0\) on the background field \(\phi_0\). For an ideal cosmic fluid, the field's Lagrangian and its Hamiltonian become, respectively, the pressure \(p(X,\phi)\), and the energy density \(\rho(X,\phi)\) of the fluid.

In ~\cite{cline} the author raised stability concerns about ~\cite{laura1}, while setting the Hubble parameter \(H=0\) ignoring the coupling of the field to gravity and its propagation on an expanding background. (Furthermore, the author approximated \(<\frac{\partial\rho}{\partial X}> = <\rho_X> \simeq \kappa\ >0\) while, in fact, inside the crystal \(<\rho_X> \simeq - \kappa/2 <0\), required by the positivity of the speed of sound squared for the phantom field. Their latter assumption on a positive \(<\rho_X> \) changed the overall sign of the total energy of perturbations, even for \(H=0\).)

The model of \cite{laura1} was constrained by three requirements: a phantom equation of state \( w < -1 \) that led to \(g'(X) <0\) therefore \(p_X <0\); a nonnegative energy density \( \rho \geq 0 \); and a positive speed of sound squared  \( c_s^2 =\frac{p_X}{\rho_X}= \frac{g'(X)}{g'(X) + 2 X g''(X)} \geq 0 \), where prime in \(g'(X)\) and \(p_X, \rho_X\) denote derivatives of the respective quantities with \(X\). Given that the phantom field requires \(w<-1\), therefore \(g'(X), p_X < 0\) then a positive speed of sound further required \(\rho_X <0\). As shown in ~\cite{laura1} all of these conditions are satisfied inside the crystal region \( X\le X_{t}= \frac{\kappa}{6\lambda}\)

Here we investigate the possibility of ghost and gradient instabilities in the field's perturbations in the context of an expanding background. We derive the equations of the scalar field and its perturbations and show that they are different from the ones presented in ~\cite{cline} due to the fact that the field is coupled to gravity and it is propagating on an accelerated expanding background, factors that were not included in ~\cite{cline}. 

First some preliminaries:
The energy-momentum tensor for a scalar field in a flat Friedmann-Robertson-Walker (FRW) universe with metric \(ds^2 = -dt^2 +a(t)^{2}dx_{(3)}^{2}\) is given by:
\begin{equation}
T_{\mu\nu} = \partial_{\mu} \phi \dfrac{\partial \mathcal{L}}{\partial (\partial^{\nu} \phi)} - g_{\mu\nu} \mathcal{L}.
\end{equation}

For a homogeneous scalar field \( \phi = \phi(t) \), the corresponding energy density \( \rho \) and pressure \( p \) of the ideal fluid are:
\begin{align}
\rho &= T^{0}_{\ 0} = \dot{\phi} \dfrac{\partial \mathcal{L}}{\partial \dot{\phi}} - \mathcal{L}, \label{eq:rho_def} \\
p &= -\dfrac{1}{3} T^{i}_{\ i} = \mathcal{L}, \label{eq:p_def}
\end{align}
where \( i \) runs over spatial indices.

The relevant Einstein equations that are sufficient for describing the ideal fluid are the Friedmann equation and the Bianchi identity.
The Friedman equation is
\begin{equation}
(\frac{\dot{a}}{a})^{2} = H^{2} = \frac{8\pi G}{3} ( \rho +\rho_{m} )
\label{eq:friedmann}
\end{equation}
where \( a(t) \) is the scale factor of the FRW metric and \(H\) the Hubble parameter, \(\rho_m,\) some matter energy density ingredient, and  \(\dot{} = d/dt\). The matter component which at present time \(t_0\) accounts for less than \(30\%\) soon dilutes away and the future cosmic evolution is solely governed by the phantom field. 

The Bianchi identity of the phantom fluid is
\begin{equation}
\dot{\rho} + 3 \frac{\dot{a}}{a} (\rho + p) = 0
\label{eq:bianchi}
\end{equation}

The non canonical conjugate momenta \(\Pi \), obtained as a derivative of \( \mathcal{L} \) with respect to \( \dot{\phi} \), is different from the kinetic momentum \( \dot{\phi} \) 
\begin{equation}
\Pi = \dfrac{\partial \mathcal{L}}{\partial \dot{\phi}} = g^{\prime}(X) \dfrac{\partial X}{\partial \dot{\phi}} =  g^{\prime}(X) \dot{\phi}= \dot{\phi} [-\kappa + \lambda \dot{\phi}^2 ],
\end{equation}
where \( g^{\prime}(X) = \dfrac{d g}{d X} \) and \( \dot{\phi} =  \pm \sqrt{2 X} \).

The Lagrangian equation of motion for the homogeneous solution is
\begin{equation}
\frac{1}{a^3}\frac{d a^{3} \Pi(X)}{dt} = \frac{\partial \mathcal{L}}{\partial \phi}
\label{eq:lagrangepi}
\end{equation}
which gives,
\begin{align}
\rho_X \ddot{\phi} + 3 H p_X \dot{\phi} + \frac{\partial V}{\partial \phi} & =0,\\
\ddot{\phi} + 3 H c_{s}^{2} \dot{\phi} + \frac{1}{\rho_X}\frac{\partial V}{\partial \phi} & = 0.
\label{eq:eomexplicit}
\end{align}

where \(c_{s}^{2}\) is the speed of sound squared which is positive within the crystal region. In the last line, Eqn.~\eqref{eq:eomexplicit}, we factored out \(\rho_X\) and used the definition of the speed of sound squared.

Comparing the above field equation, Eqn.~\eqref{eq:eomexplicit}, with Eqn.2 of ~\cite{cline}, reveals that coupling to gravity, therefore the Hubble drag term, where missing from ~\cite{cline}. 
There is a general misconception that \(\rho_X < 0\) produces instabilities. This could be true and present a problem if this term produced a negative speed of sound squared, i.e. only for \(p_X >0\). However, as can be seen from Eqn.~\eqref{eq:eomexplicit}, in the present case the only influence a negative \(\rho_X\) term, (given that \(c_{s}^{2} >0\)), has in the field equation of motion is in inverting the potential energy term in the homogeneous mode and perturbations equations (below). 
Hence, the speed of sound squared multiplying the Hubble term plays a crucial role as this term dominates the equation. The potential term is insignificant compared to the Hubble drag term, therefore it doesn't control the dynamics of the field. In short, stability of the model depends on the ratio of \(p_X\) and \(\rho_X\). For as long as their ratio equal to the speed of sound squared \(c_{s}^{2} =p_X /\rho_X >0\) remains positive, the system is stable. 

Note that for a phantom field, the Hubble parameter grows with time, therefore the Hubble drag term quickly dominates over the potential term \(V(\phi)\) making it irrelevant even faster in the dynamics of the field.

The growth of \(H, a(t),\rho\) can be found by solving the Friedmann equation and the Bianchi identity for a phantom field with an equation of state such that \( (1+w)<0\) that give the following expressions ~\cite{caldwell} for the scalar factor,
\begin{equation}
a(t) = a(t_{0}) [-w + (1+w) t/t_{0}]^{\frac{2}{3(1+w)}}
\label{eq:scalefactor}
\end{equation}
with \(t_0\) the present time.
The Hubble parameter,
\begin{equation}
    H \simeq [\frac{a(t)}{a(t_{0})}] ^ {\frac{-3}{2(1+w)}}
\label{eq:hubble}
\end{equation}
and the the energy density of the homogeneous field
\begin{equation}
    \rho \simeq  ( \frac{a(t)}{a(t_0)})^{-3(1+w)}
    \label{eq:energydensity}
\end{equation}

In the above, factors of \(8\pi G/3\) are ignored by setting the reduced Planck mass to one.

In deriving and in attempting to get an approximate analytic solution for the perturbation equations below, we will for simplicity ignore the metric perturbations (set \(\delta g_{\mu\nu} =0\)) and focus solely on the scalar field. in this case the perturbations equation becomes

\begin{equation}
    \rho_{X} \ddot{\delta\phi} - \frac{p_{X}}{a^2}\delta\phi_{,ii} + (3 H \rho_{X} + 6 \lambda\dot{\phi}\ddot{\phi}) \dot{\delta\phi} + \frac{\partial^{2}V}{\partial \phi^2}\delta\phi = 0
\end{equation}
which, when dividing by \(\rho_X\) and recalling the definition of the speed of sound and the fact that \(\rho_X, p_X\) are both negative, can be further simplified
\begin{equation}
   \ddot{\delta\phi_{k}}  + \frac{c_{s}^{2} k^{2}}{a^2}\delta\phi_{k} + 3 H \dot{\delta\phi_{k}} - \frac{6 \lambda \ddot{\phi}\dot{\phi}}{|\rho_X|} \dot{\delta\phi_{k}} - \frac{m^2}{|\rho_X|}\delta\phi_{k} =0 
   \label{eq:perturbations}
\end{equation}
Notice that in the above, we replace \(V_{\phi\phi} \simeq m^2\) and assumed the free plane wave for the spatial part \(\delta\phi_k \simeq f(t) e^{ikx}\). It can be seen that the speed of sound for the perturbations propagation is the same as that of the background field (the fourth \(\lambda\) term can be ignored relative to the Hubble drag and it averages to zero), and it is positive.
(If we included metric perturbations, the above equations become very complicated and can only be solved numerically. We will report the results of the numerical analysis in an upcoming paper. To get an idea, we could use the synchronous gauge in which case the right hand side would have a factor of the trace of perturbations \(h \dot{\phi}\) instead of 0, or use the Mukhanov variables in conformal time for estimating the CMB spectrum.)

The last two terms in Eqn.\eqref{eq:perturbations} are insignificant relative to the ever increasing Hubble drag term. However, an approximate solution can be obtained even with those terms included, as follows. The avid reader can identify this equation with the damped pendulum. For the time dependent part \(f(t)\) of \(\delta\phi_k\), assume \(f(t)\simeq Exp[-\beta t]\) and replace it in Eqn.~\eqref{eq:perturbations}, to get the following solutions
\begin{equation}
    \beta_{1,2} = \frac{(3 H - \frac{6\lambda\ddot{\phi}\dot{\phi}}{|\rho_X|}) \pm \sqrt{(3 H - \frac{6\lambda \ddot{\phi}\dot{\phi}}{|\rho_X|})^2 + 4 (\frac{m^2}{|\rho_X|} - \frac{k^2 c_{s}^{2}}{a^2})}}{2} >0
    \label{eq:solutions}
\end{equation}

For \(\beta <0\) we would have growth of perturbations therefore instability. In the present model, there is only one exception when this might happen: that is, the minus sign solution for \(\beta\) in Eqn.~\eqref{eq:solutions} for \textit{only} a small range in the longest wavelength modes \(\bar{\lambda} \simeq 2\pi/k \rightarrow\infty\) corresponding to co moving wavenumbers \(0\leq k^2 \le \frac{2m^2}{\kappa}\). Yet, even for these longest super horizon wavelengths, the unstable solution which goes as: \(e^{\frac{(m^{2}/\kappa - k^2/a^2)}{3H} t}\), is suppressed and driven to zero quickly by the factor of \(3H\) in the denominator which overwhelmingly dominates the mass term in the nominator. Therefore, it can be ignored.

With the above caveat, it can be seen that approximately for super-horizon modes \(k^2 \le 9/4 (Ha)^2\) we always have \(\beta_{1,2} >0\) solutions therefore those modes are overdamped (decaying) \(\delta\phi_k \simeq e^{- \beta t}\). For sub-horizon modes with \(k^2 \ge 9/4 (Ha)^2\) we can write \(2\beta_{1,2} \simeq 3H \pm i\Gamma\), with \(\Gamma\simeq \sqrt{\frac{c_{s}^2 k^2}{a^2} -(3H)^2}\), where we dropped the \(\lambda, m^2\) terms as sub-leading to avoid clutter.(They can easily be restored back in the definition of \(\beta_{1,2},\Gamma\).) For sub-horizon perturbations we find an under-damped solution, that is an oscillatory behavior modulated by a decaying function, \(\delta\phi_k \simeq A e^{-3Ht} \cos(\Gamma t)\),
(or \(\sin(\Gamma t)\) depending on the matching of oscillating to decaying modes at the horizon \(k \simeq Ha\).)
The total energy density, up to second order in perturbations is given by
\begin{equation}
    \delta ^{(2)}\mathcal{H} =  \rho_X [ \dot{\delta\phi_k}^2 + ( \frac{k^2 c_{s}^2}{a^2} - \frac{m^2}{|\rho_X|} ) \delta\phi_k^2 ]
    \label{eq:energyperturb}
\end{equation}

As noted before, \(\rho_X <0\) and \(c_{s}^2 = p_{X}/\rho_X >0\)  , therefore when we replace it with the absolute value in the mass term \(V_{\phi,\phi}=m^2\), we pick a minus sign.
Ultimately, the scale factor, Eqn.~\eqref{eq:scalefactor} diverges in a finite time \(t\simeq t_{o} w/(1+w)\). According to the Lagrangian equation of motion, Eqn.~\eqref{eq:lagrangepi}, the conjugate momentum of the field is driven to zero as \(\Pi(X) \simeq N/a^3\), with \(N=constant\), which implies \(\Pi(X) \simeq \frac{N}{a(t_0)^3} [-w +(1+w) t/t_{0}]^{\frac{-2}{1+w}}\) is dragged to zero during the same time interval \(t\simeq t_{0} w/(1+w)\) as the scale factor diverges and the universe goes through the Big Rip. Replacing the approximate solutions in Eqn.~\eqref{eq:energyperturb}, and noting that for the crystal on average \(\rho_X \simeq -\kappa/2\) we obtain for the rate of change over time of the perturbations energy

\begin{equation}
\begin{split}
& dE/dt = 2 \dot{\delta\phi_k}  [\rho_X \ddot{\delta\phi_k} + (\frac{k^2 p_X}{a^2} + m^2)\delta\phi_k]\\
& =-2 \dot{\delta\phi_k}^2 (3H - 6\lambda \dot{\phi}\ddot{\phi}) \simeq -6 H \dot{\delta\phi_k}^2 <0 
\end{split}
\end{equation}

where \(E\) is the energy density of the Hamiltonian up to second order of Eqn.~\eqref{eq:energyperturb}.
Since \(\dot{\delta\phi_k}^2 >0\) and \(dE/dt <0\), the perturbation system is losing energy in the under-damped sub-horizon oscillations or overdamped super horizon modes driven to zero by the Hubble drag term. At the interface of the crystal, \(X=X_t, \rho_X =0\) this energy is exactly zero. In contrast the energy of the homogeneous mode \(\rho\) grows with time as given by Eqn.~\eqref{eq:energydensity}.

The approximate analytical results presented here should be treated with caution. Besides ignoring metric perturbations, we also treated the Hubble parameter as nearly a constant in obtaining the approximate perturbation field solutions. In reality, the Hubble parameter is driven by the energy of the phantom field and the exact field solutions depend on the time dependence of the Hubble parameter. The two are coupled to each other via Einstein's equations. However, these equations can only be solved numerically. Although the stability of perturbations will not change, it will be interesting to find out how the metric and the field dynamics evolve as we approach the Big Rip, and as the field is dragged to an attractor fixed point \(\Pi(X) =0\). We will report these numerical results in an upcoming paper.

\noindent\textit{Acknowledgment:}
L.Mersini-Houghton is grateful to the Klinsberg foundation for their support.

\end{document}